\documentclass[aps,prl,superscriptaddress,preprintnumbers,
showpacs,legalpaper,twoside,twocolumn]{revtex4}

\usepackage{bm}
\usepackage{graphicx}
\usepackage{amsfonts}
\usepackage{amsmath}
\usepackage{amssymb}

\begin{document}

\title{Quantum localization in bilayer Heisenberg antiferromagnets
with site dilution}

\author{Tommaso Roscilde}
\affiliation{Department of Physics and Astronomy, University of Southern
California, Los Angeles, CA 90089-0484}
\author{Stephan Haas}
\affiliation{Department of Physics and Astronomy, University of Southern
California, Los Angeles, CA 90089-0484}

\pacs{75.10.Jm, 75.10.Nr, 75.40.Cx, 64.60.Ak}

\begin{abstract}

The field-induced antiferromagnetic ordering in systems of weakly
coupled $S=1/2$ dimers at zero temperature can be described as a Bose-Einstein 
condensation of triplet quasiparticles (singlet quasiholes) in the ground 
state. For the case of a Heisenberg bilayer, it is here shown 
how the above picture is altered in the presence 
of site dilution of the magnetic lattice. Geometric randomness leads to 
quantum localization of the quasiparticles/quasiholes and to 
an extended Bose-glass
phase in a realistic disordered model. This localization phenomenon
drives the system towards a quantum-disordered phase well before 
the classical geometric percolation threshold is reached.
\end{abstract}
\maketitle

 The introduction of randomess in strongly correlated
fermionic and bosonic systems on a lattice offers a large showcase 
of novel quantum behavior \cite{randomsingletphase,Fisheretal89,
ShenderK91}. In quantum spin systems 
with an ordered ground state, the generic effect of geometric 
disorder is an enhancement of local quantum fluctuations
through the reduction of the connectivity of the lattice
\cite{Sandvik02},
which may lead to quantum disordered phases \cite{Yuetal05}. 
On the other hand, 
in spin systems with a gapped quantum-disordered ground state,
geometric randomness can induce local magnetic moments
giving rise to long-range order through an \emph{order-by-disorder} 
mechanism \cite{ShenderK91}. In the case of superfluid interacting 
bosons, the presence of an increasing amount of disorder can destroy
superfluidity through quantum localization, inducing
a disordered, yet gapless \emph{Bose-glass} phase 
with a finite compressibility \cite{Fisheretal89}.
 
Recently, the strong connection between lattice bosons and 
quantum spin models has been pointed out in the context of
unfrustrated $S=1/2$ weakly-coupled dimer systems. 
For sufficiently weak inter-dimer couplings, 
the spin system has a singlet ground state with 
a finite gap to triplet excitations. The application of a 
uniform magnetic field leads to a lowering of the energy
of the triplet excitations aligned with the field, until,
at a lower critical field $h^{(0)}_{c1}$, the ground state develops
a finite magnetization. Within a bosonic mapping
that neglects the other two triplets, bosonic hardcore triplet 
quasi-particles (TQPs) populate the ground state forming 
a bosonic condensate \cite{Rice2002}.
Off-diagonal long-range order corresponds
to the appearence of a finite staggered magnetization in the plane
perpendicular to the field. When the magnetization is half-saturated, 
the relevant degrees of freedom of the condensate
become hardcore singlet quasi-holes (SQHs) in the triplet sea, 
which disappear completely at an upper critical field $h^{(0)}_{c2}$
fully polarizing the system.
For $h<h^{(0)}_{c1}$  ($h>h^{(0)}_{c2}$) the system is therefore 
a band insulator, with an empty (full) band of TQPs. 
The validity of the description of the ordered phase as 
a Bose-Einstein condensate has been extensively 
investigated both theoretically and experimentally 
\cite{Cavadinietal03,Jaimeetal04,Nikuni00}.
 
 By virtue of the bosonic picture, it is intriguing
to envision the possibility of a Bose-glass phase for 
weakly coupled dimer systems in a magnetic 
field upon introduction of lattice disorder 
\cite{ShindoT04}. This opens up
the appealing perspective of unambiguously realizing 
such a phase in a quantum spin system, profiting of the 
high level of control with which disorder can be 
introduced in a magnetic lattice \cite{Vajketal02}. 
The focus of this Letter is the extent and the major 
signatures of such a phase in a realistic model.

In what follows we investigate the $S=1/2$ Heisenberg 
antiferromagnet in a magnetic field and with site dilution, 
whose Hamiltonian reads:
\begin{eqnarray}
{\cal H} &=& J' \sum_{\langle ij\rangle} \sum_{\alpha=1, 2}
\epsilon_{i,\alpha} \epsilon_{j,\alpha} 
 {\bm S}_{i,\alpha}\cdot{\bm S}_{j,\alpha}  \nonumber \\
&+& J \sum_{i} \epsilon_{i,1} \epsilon_{j,2} 
 {\bm S}_{i,1}\cdot{\bm S}_{i,2}
 - g\mu_B H \sum_{i,\alpha} \epsilon_{i,\alpha} S^z_{i,\alpha}~.
 \label{e.hamilton}
\end{eqnarray}
Here the index $i$ runs over the sites of a square lattice, 
$\langle ij\rangle$ are pairs of nearest neighbors on the 
square lattice, and $\alpha$ is the layer index.
$J$ is the \emph{inter}layer coupling and $J'$ 
the \emph{intra}layer one.
The variables $\epsilon_{i,\alpha}$ take the values 0 or 1
with probability $p$ and $1-p$ respectively, $p$ being
the concentration of non-magnetic sites. Hereafter we will
express the field in reduced units $h = g\mu_B H/J$.
We have investigated the above Hamiltonian making use
of Stochastic Series Expansion quantum Monte Carlo
based on the directed-loop algorithm \cite{SyljuasenS02},
and considering $L\times L\times 2$ lattices with $L$ 
up to $40$. The $T=0$ limit is reached through 
a $\beta$-doubling approach \cite{Sandvik02}. Disorder
averaging is  performed by using typically $200$ 
different disorder realizations. 

\begin{figure}[h]
\begin{center}
\includegraphics[bbllx=60pt,bblly=50pt,bburx=510pt,bbury=450pt,%
     width=55mm,angle=0]{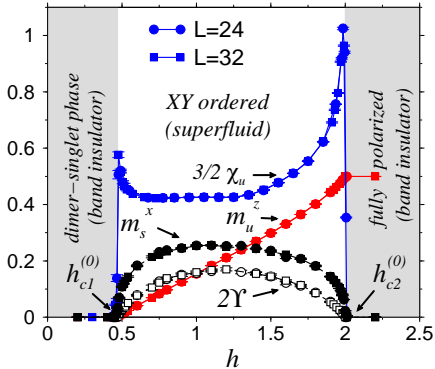} 
\caption{Field dependence of the uniform magnetization
$m_u^{z}$, transverse staggered magnetization $m_s^{x}$,
and superfluid density $\Upsilon$ at $T=0$ for the 
bilayer antiferromagnet with $J/J'=4$ in the clean limit
($p=0$).}
\label{f.clean}
\vskip -.3cm
\end{center}
\end{figure}

In the clean limit ($p=0$) and at zero field, the 
Hamiltonian of Eq.(\ref{e.hamilton}) is in
a gapped dimer-singlet phase for
$J/J' \gtrsim 2.5$ \cite{SandvikS94}. Hereafter we will 
specialize to the case of a bilayer with $J/J' = 4$, namely 
well inside the dimer-singlet phase. 
 As discussed in the introduction, the clean system
with large $J/J'$ ratio can be well described by an effective 
model of hardcore TQPs which hop from dimer to dimer
with amplitude $J'/2$, nearest-neighbor repulsion $J'/2$,
and chemical potential $J(h-1)$ controlled by 
the field \cite{GiamarchiT99}. Increasing the field 
leads to the closing the triplet gap, accompanied by 
the condensation of TQPs. 
We have investigated this scenario 
numerically, by studying the field evolution of the
uniform magnetization $m_u^{z} = \langle S_i^{z} \rangle$,
the uniform susceptibility $\chi_u = dm_u^{z}/dh$,
the staggered magnetization 
$m_s^{x} = 1/4 \sum_{\alpha\beta} 
\sqrt{ (-1)^{L/2+\alpha+\beta} \langle S_{i,\alpha}^{x} 
S^{x}_{i+L/2,\beta} \rangle}$,
and the superfluid density $\Upsilon = T/2 ~\langle W_x^2 + W_y^2 \rangle$,
where $W_{x(y)}$ is the worldline winding number 
in the $x$($y$) direction.
As shown in Fig. \ref{f.clean}, a field-induced ordered 
phase with $m_s^{x}, \Upsilon \neq 0$ 
is clearly observed between $h^{(0)}_{c1} = 0.47(1)$ and 
$h^{(0)}_{c2} = 2$,
in between a dimer-singlet phase ($h<h^{(0)}_{c1}$) and a fully
polarized phase ($h<h^{(0)}_{c2}$).

 We now turn to the case of the doped system, $p\neq 0$. 
From the purely geometric point of view, the lattice
undergoes a percolation transition at $p^{*} = 0.5244(2)$
which we estimated through classical Monte Carlo calculations
based on a recently proposed algorithm \cite{NewmanZ01}.

\begin{figure}[h]
\begin{center}
\includegraphics[bbllx=50pt,bblly=50pt,bburx=600pt,bbury=450pt,%
     width=70mm,angle=0]{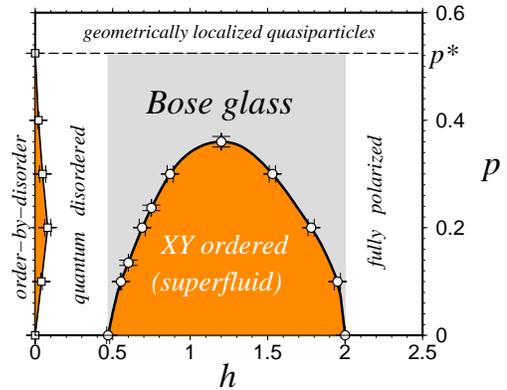} 
\caption{Ground-state phase diagram of 
the site-diluted bilayer Heisenberg antiferromagnet 
with $J/J'=4$.}
\label{f.phdiagr}
\end{center}
\end{figure}

The magnetic behavior in turn displays a rich 
sequence of phases well below the percolation
transition.
Fig. \ref{f.phdiagr} shows the zero-temperature 
phase diagram, where boundary lines between ordered and
disordered phases have been estimated
through the linear scaling of the disorder-averaged 
correlation length, $\xi^{xx(yy)}\sim L$. 


\begin{figure}[h!]
\begin{center}
\includegraphics[bbllx=0pt,bblly=50pt,bburx=580pt,bbury=520pt,%
     width=85mm,angle=0]{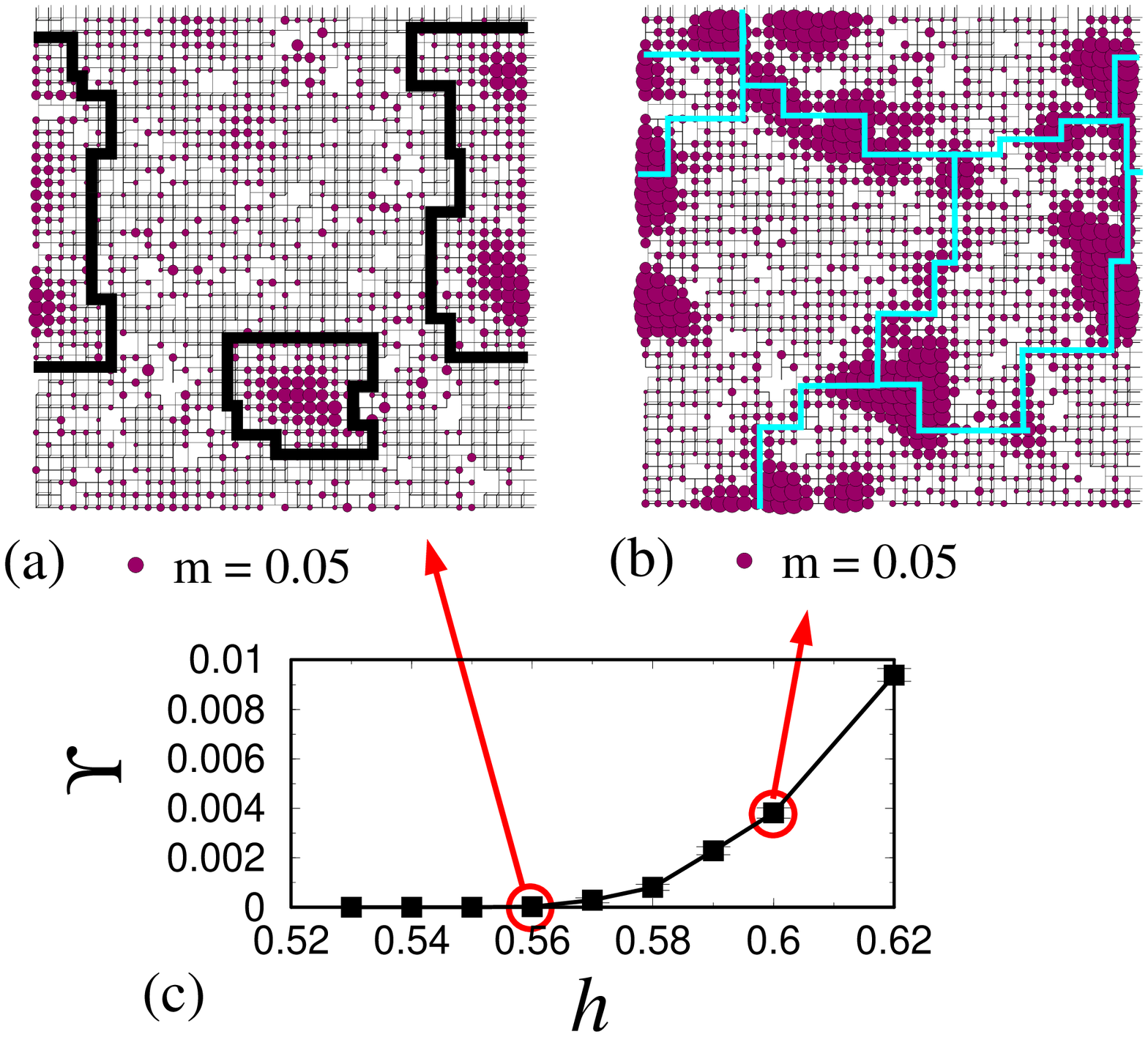} 
\caption{(a)-(b): Real-space images of the dimer
magnetization $m_i=\langle S^{z}_{i,1} + S^{z}_{i,2} \rangle$ 
on intact dimers in a 40x40x2 bilayer with $J/J'=4$, 
dilution $p=0.1$ and at
inverse temperature $\beta J = 256$, for $h = 0.56$
(a) and $h = 0.6$ (b). The radius of the dots is proportional
to the dimer magnetization.
The magnetization of unpaired spins is omitted for 
clarity. The most visible localized states are 
highlighted in (a), while the backbone of the percolating
magnetized network is highlighted in (b).
(c): Superfluid density as a function of the field for the 
specific sample considered. }
\label{f.BGtoSF}
\vskip -.8cm
\end{center}
\end{figure}  

In what follows, we will discuss the phase diagram 
making use of both the bosonic and the magnetic language.
Strictly speaking, the conventional bosonic mapping
onto hardcore TQPs breaks down in presence of impurities,
since some of the spins become \emph{unpaired},
as they miss their partner on the neighboring layer.
A bosonic mapping is however still possible in 
which bosons correspond to $|\uparrow\uparrow\rangle$
states on intact dimers and to $|\uparrow\rangle$
states on unpaired spins. Site dilution of the 
spin model reflects in correlated diagonal and 
off-diagonal disorder of the bosonic model \cite{Roscildeunp}. 

In the dimer-singlet phase the introduction of vacancies
in the magnetic lattice leads to the 
appearence of local $S=1/2$ moments. It is straightforward
to identify these moments with unpaired spins, but in fact 
they occupy a larger volume $\sim\xi_0^{D}$ where $\xi_0$ is 
the correlation length in the clean limit $p=0$ \cite{Sandviketal96},
thus spreading also over intact dimers. 
Those moments are coupled through an effective
unfrustrated antiferromagnetic network, which can sustain
long-range order at $T=0$. This gives rise to 
an \emph{order-by-disorder} phenomenon, which is 
expected to persist for any doping concentration 
up to percolation $p^{*}$. 
The network of local moments has
a broad distribution of effective couplings, 
which scale exponentially with the inter-moment 
distance $r$ \cite{SigristF96}, $J_{\rm eff} \sim \exp[-r/\xi_0]$. 
The application of a moderate field ($h<h_{c1}^{(0)}$) can easily
destroy the antiferromagnetic order by polarizing 
a majority of the effective magnetic moments \cite{Mikeskaetal04},
bringing the system into a \emph{partially
polarized quantum-disordered} phase. 
In particular, those dimers which are immediately close to 
an unpaired spin (roughly within a distance of $\xi_0$) 
are also partially polarized, even if the field is 
lower than the dimer gap and the bilayer gap. 
In the bosonic picture, this means that TQPs 
are present in the ground state, localized around unpaired spins 
and only weakly penetrating clean regions of intact dimers, 
which mostly remain unmagnetized. They form a
``gossamer" superfluid network with exponentially 
suppressed links for small fields, but for larger particle 
numbers this network turns into an insulating one due to 
the complete filling of some of its islands. 

Upon increasing the field beyond $h^{(0)}_{c1}$, the 
picture of TQPs exponentially localized around vacancies
start to change substantially. Continuous regions
of intact dimers, which occur with probability exponentially
small in the size of the region, begin to respond to 
a field similarly to what would happen in the clean case.
Some of the intact dimers with a lower local coordination, 
nonetheless, have a higher local gap which a
field $h\approx h^{(0)}_{c1}$ is not able to close, so that
they remain non-magnetized up to higher fields.  
In the bosonic picture, this means that TQPs begin to 
appear in the \emph{bulk} of the clean regions, but 
they are exponentially localized due to disorder.
This picture clearly corresponds to that of a Bose-glass
phase \cite{Fisheretal89}. Such a phase extends up to 
a critical field $h_{c1}$ at which 
the increased population of bosons, together with
the hardcore repulsive interaction, leads to a 
localization-delocalization transition and to 
superfluidity, as shown in Fig. \ref{f.BGtoSF}.
There the density distribution of the TQPs
is imaged by the dimer magnetization 
$m_i=\langle S^{z}_{i,1} + S^{z}_{i,2} \rangle$.
As the figure shows, superfluidity occurs via quantum
percolation of the \emph{collective} bosonic state throughout
the lattice, while single particle states would all be 
exponentially localized in 2D.

 From the numerical data, we observe that the Bose-glass
phase extends over a field region that increases with the
dilution $p$. In particular, beyond an upper critical 
dilution, $p_c \approx 0.36$, no field-induced order survives.
This happens well below the lattice percolation threshold 
$p^{*}$. Therefore there is quite a large region of doping
completely dominated by quantum disorder.

 As the field is increased even further for $p < p_c$, 
 singlet quasi-holes (SQHs) become the relevant
degrees of freedom. They undergo a similar 
superfluid-to-Bose-glass transition at a field $h_{c2}$ 
without major alterations with respect to the TQP
case. In particular the magnetization does not reach full 
saturation until the clean upper critical field $h^{(0)}_{c2}$,
given that (exponentially rare) large clean areas 
cannot be fully polarized before $h^{(0)}_{c2}$ is reached.
\begin{figure}[h]
\begin{center}
\includegraphics[bbllx=32pt,bblly=44pt,bburx=580pt,bbury=395pt,%
     width=80mm,angle=0]{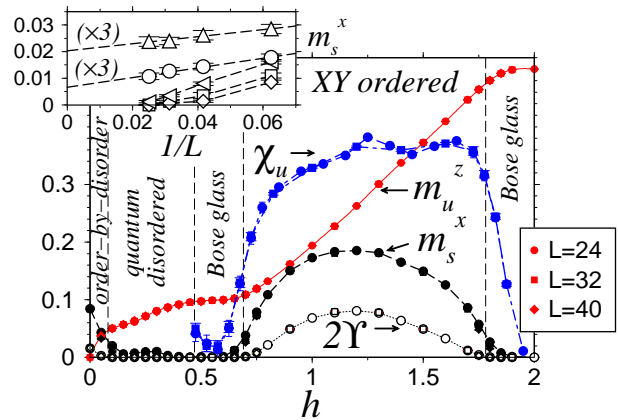} 
\caption{Zero-temperature field scan in 
the site-diluted bilayer Heisenberg antiferromagnet 
with $J/J'=4$ and $p=0.2$. Inset: scaling of the order parameter
$m_s^x$ in the different phases of the system; from top
to bottom $h=0.75$ (XY ordered), $0.05$ (order-by-disorder), $1.85$
(high-field Bose glass), $0.35$ (quantum disordered), $0.6$
(low-field Bose glass). }
\vskip -.8cm
\label{f.hscan}
\end{center}
\end{figure} 

 A typical field scan through the phase diagram is shown
in Fig. \ref{f.hscan} for a dilution $p=0.2$. 
The field is 
seen to quench the order-by-disorder spontaneous transverse
magnetization and to drive the system towards a partially
polarized state \emph{without} fully polarizing the 
local free moments, whose magnetization continues
to grow even beyond the transition. Eventually
for large enough fields all the local moments get fully
polarized by the field, leading to a magnetization
plateau at $m = p S = p/2$ \cite{footnote}. 
In the Bose-glass phase
the uniform magnetization is seen to change concavity 
and to start to grow slowly, as the clean regions
are gradually magnetized. A simple analysis can be made
based on a ``local-gap'' model, for which the overall 
magnetization is the sum of the magnetizations of uncorrelated clusters 
of intact dimers with different local gaps \cite{Roscildeunp}. 
The result is that the growth over the plateau value 
is \emph{exponentially activated},
$m_u^{z} - pS \sim \exp\left[-c\left(h-h_{c1}^{(0)}\right)^{-1}\right]$,
where the exponential behavior comes from the 
exponential tail of the size distribution of clean regions,
and it is a direct evidence of the dominant behavior
of \emph{rare events} in the Bose-glass phase.  Similarly, 
an exponential saturation is observed for $h\to h_{c2}^{(0)}$ 
in the upper Bose-glass region.

\begin{figure}[h]
\begin{center}
\includegraphics[bbllx=30pt,bblly=140pt,bburx=590pt,bbury=480pt,%
     width=75mm,angle=0]{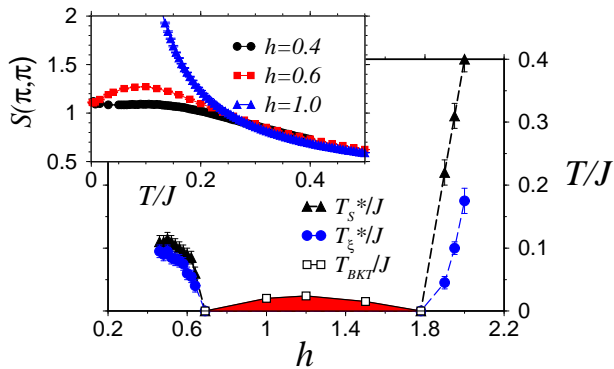} 
\caption{Finite-temperature phase diagram for the 
site-diluted bilayer Heisenberg antiferromagnet 
with $J/J'=4$ and dilution $p=0.2$. In the inset: temperature
dependence of the transverse structure factor for the 
same model with $L=32$.}
\label{f.finiteT}
\vskip -.5cm
\end{center}
\end{figure}

 Let us now turn to the finite-temperature signatures of each 
phase. The zero-temperature XY ordered phase turns into a
quasi-long-range ordered phase at finite $T$ up to 
a Berezinskii-Kosterlitz-Thouless (BKT) transition 
temperature $T_{_{\rm BKT}}$, at which the superfluid density
vanishes and the transverse
structure factor $S(\pi,\pi) = (1/N)\sum_{ij,\alpha\beta} 
(-1)^{i+j+\alpha+\beta} \langle S_{i,\alpha}^{x(y)}
S_{j,\beta}^{x(y)}\rangle$ becomes finite. 
We estimate the transition temperature $T_{_{\rm BKT}}$ through 
BKT scaling of the superfluid density and of the transverse
structure factor \cite{Cuccolietal03}. The transition temperature 
as a function of the field is shown in Fig. \ref{f.finiteT}.
Due to disorder and quantum fluctuations, $T_{_{\rm BKT}}$
is almost two order of magnitudes lower
than the interlayer exchange coupling $J$, and its slightly
asymmetric field dependence mimics that of the $T=0$
order parameter. 

 Turning on the temperature from a $T=0$ disordered phase,
 on the other hand,  
 we observe interesting differences between the quantum
 disordered phase for $h<h_{c1}^{0}$ and the Bose-glass phase
 for $h_{c1}^{0}<h<h_{c1}$ and $h_{c2}<h<h_{c2}^{0}$. 
 In the clean limit $p=0$, 
 the gapped disordered phases for $h<h_{c1}^{0}$ and
 $h>h_{c2}^{0}$ show the common feature of 
 \emph{thermal activation},
 with transverse structure factor $S(\pi,\pi)$, 
 transverse correlation length $\xi^{xx(yy)}$, and uniform
 magnetization $m_u^z$ increasing as $T$ grows above 0.
 Such a behavior is easily understood: upon 
 increasing the temperature, the system is able 
 to explore more correlated 
 states with finite magnetization sitting beyond the energy gap.
 In the presence of site dilution, 
 the temperature activation of correlations 
 turns into a slow decrease as $T$ increases
 in the low-field regime dominated by the unpaired spins
 (see inset of Fig. \ref{f.finiteT}). For high enough 
 fields, and in particular \emph{in the Bose-glass phase}, 
 the temperature activation of correlations is restored
 up to a temperature $T^{*}_{S}$ for $S(\pi,\pi)$ 
 and $T^{*}_{\xi}$ for $\xi$ at which both quantities 
 hit a maximum, showing then a crossover to a decreasing 
 dependence typical of a conventional paramagnetic phase.
 In contrast to the clean disordered phases, 
 the Bose-glass phase is gapless. Nonetheless some clusters
 of intact dimers resist being magnetized 
 for $h_{c1}^{(0)}<h<h_{c1}$ or they are fully polarized 
 already for $h_{c2} <h<h_{c2}^{(0)}$. 
 This means that the observed temperature activation 
 in that field range can be associated with those
 portions of the system visiting states with higher 
 transverse correlations, linked to the
 presence of bosonic TQPs/SQHs. In the 
 bosonic language, this can be suggestively pictured
 as a partial delocalization of the trapped quasiparticles
 into those regions, assisted by thermal fluctuations. 
    
  Our findings are directly relevant to experimental 
  investigations of site-diluted weakly coupled dimer 
  systems in a magnetic field. The particular case
  of an Heisenberg antiferromagnet on coupled bilayers 
  is realized by BaCuSi$_2$O$_6$ \cite{Jaimeetal04},
  in which doping of the magnetic Cu$^{2+}$ ions with non 
  magnetic Zn$^{2+}$ and Mg$^{2+}$ can lead to site dilution 
  of the magnetic lattice. The picture of quantum localization 
  of bosonic quasiparticles applies nonetheless to other  
  spin gap systems with different geometries, such as 
  Sr$_2$Cu(BO$_3$)$_2$ \cite{Sebastianetal05} or Tl(K)CuCl$_3$
  \cite{Cavadinietal03}.
  The quantities we have
  investigated clearly show a characteristic sequence of phases, 
  with disorder suppression of magnetism and onset of a 
  novel Bose-glass phase, and they are directly accessible
  to magnetometry and neutron scattering experiments.
  
  We acknowledge fruitful discussions with C. Batista, M. Jaime, N. Laflorencie,
  B. Normand, N. Prokof'ev, H. Saleur, and T. Vojta, and we particularly
  thank O. Nohadani and S. Wessel for carefully reading 
  our manuscript. This work is supported by DOE. Computational 
  facilities have been generously provided by the HPC Center at USC.


\begin{thebibliography}{99}
\bibitem{footnote} For the specific model considered the 
magnetization reaches the value $p/2$ only beyond
$h^{(0)}_{c1}$, which means that the plateau shrinks to 
an horizontal inflection point and the overall quantum
disordered phase for $h<h^{(0)}_{c1}$ is gapless. 
For a larger $J/J'$ an extended 
magnetization plateau appears below $h^{(0)}_{c1}$,
so that the disordered phase for $h\lesssim h^{(0)}_{c1}$
becomes gapped \cite{Roscildeunp}.
\end{thebibliography}
\end{document}